# Overcoming Dynamic Range Limitations in Computational Hyperspectral Imaging

*Qianqian Yu*[1], *Jinhang Lin*[1], *Zhihai Yao*[2], *Shuo Wang*[2], *Chaofan Weng*[3], *and Xiang Hao*[1,4,*]

[1] State Key Laboratory of Extreme Photonics and Instrumentation, College of Optical Science and Engineering, Zhejiang University, Hangzhou 310027, China.

[2] Jilin Qiushi Spectral Data Technology Co., Ltd., Changchun 130117, China.

[3] Hangzhou National Camera Testing Technology Co., Ltd., Hangzhou 310013, China.

[4] ZJU-Hangzhou Global Scientific and Technological Innovation Center, Hangzhou 311200, China.

*haox@zju.edu.cn

KEYWORDS: computational hyperspectral imaging, high-dynamic-range imaging, multi-exposure fusion, physics-informed neural network, broadband spectral encoding

**ABSTRACT.** Computational hyperspectral imaging offers compact spectral sensing, yet imaging high-dynamic-range (HDR) scenes remains challenging. Accurate spectral reconstruction requires encoded measurements to remain within the calibrated linear-response range and above the noise floor. Due to unequal spectral responsivities across the encoding mask, individual channels reach their reliability limits at vastly different incident signal levels. Measurement reliability therefore depends on both exposure time and encoding channel, making it challenging for a single exposure to preserve spectral information in highlights and shadows simultaneously. Here, we demonstrate a physics-guided multi-exposure fusion framework that operates directly in the encoded measurement domain. An analytical reliability prior and learned spatially varying, channel-specific adjustments jointly determine adaptive fusion weights. Measurement-domain spectral-angle supervision further constrains relative channel responses for downstream spectral inversion. The set-wise formulation accommodates varying numbers and configurations of input exposures. We further implement the framework on a 9-channel broadband-encoded stochastic hyperspectral camera, incorporating full-field sensor-response characterization and channel-aware reliability gating. Simulated HDR benchmarks show lower reconstruction errors than representative fusion methods, while controlled spectral measurements and practical HDR scenes confirm improved spectral fidelity and extended usable dynamic range. Measurements using an optical-density target further establish a 16.0-dB increase in spectral reconstruction dynamic range relative to the best fixed single exposure.

## 1. Introduction

Hyperspectral imaging (HSI) captures spatially resolved spectral information beyond color imaging, enabling applications[1–5] in remote sensing, agriculture, and material analysis. Snapshot computational HSI[6–8] replaces bulky, scanning-based spectral cameras by coupling broadband spectral modulation structures directly with complementary metal-oxide-semiconductor (CMOS) image sensors[9–11]. This architecture compresses a high-dimensional spatial–spectral cube into a single two-dimensional frame, which is subsequently reconstructed using compressive sensing algorithms[12–15] or deep neural networks[16–18].

However, capturing high-dynamic-range (HDR) scenes with computational HSI remains challenging because spectral reconstruction depends on the quality of the underlying measurements. Short exposures preserve bright-region measurements but suffer from drastically reduced signal-to-noise ratios in shadow regions. Conversely, long exposures improve weak signals but introduce severe nonlinearity or sensor saturation in highlight regions. This trade-off is further complicated by unequal spectral responsivities across the encoding mask[19], which cause individual channels to reach their reliability limits at vastly different incident signal levels. Measurement reliability therefore depends on both exposure and encoding channel. Unreliable measurements in either extreme propagate through the spectral inversion process, resulting in amplified noise and heavily distorted spectral shapes[20].

Multi-exposure acquisition[21–27] is widely used in HDR imaging to combine complementary information from different exposures, but existing methods are predominantly optimized for visual display quality in final RGB images. Recent RAW-domain fusion techniques[28,29] attempt to retain more sensor information by bypassing standard image signal processors. However, they are still designed to produce a final visual image rather than an intermediate measurement for downstream algorithmic inversion. More recently, multi-exposure HDR reconstruction has been investigated in spatially amplitude-coded snapshot spectral imaging[30], whereas its application to broadband spectrally encoded computational HSI remains less explored. In the latter, arbitrary changes to the relative responses across encoding channels can distort the underlying spatial–spectral multiplexing and compromise subsequent spectral reconstruction.

To address this challenge, we develop a physics-guided multi-exposure fusion framework to recover HDR encoded measurements before spectral reconstruction. The framework performs reliability-aware fusion directly in the encoded measurement domain, where an analytical physical prior and a learned adjustment jointly determine spatially and channel-wise varying fusion weights across exposures. Measurement-domain spectral angle supervision further constrains the relative responses across encoding channels to support accurate spectral inversion. The shared, set-wise formulation supports flexible numbers and configurations of input exposures. We validate the framework through simulations and experiments on a 9-channel broadband encoded hyperspectral camera, incorporating full-field sensor-response characterization and channel-aware reliability gating for real-system implementation. The results demonstrate improved measurement recovery and spectral fidelity, together with an extended usable dynamic range for hyperspectral reconstruction in practical HDR scenes.

## 2. Principle

The workflow of computational HSI is illustrated in Figure 1(a). A broadband modulation layer and CMOS image sensor encode the incident spectrum into 9-channel measurements. Each encoding vector is subsequently mapped to a 31-band spectrum spanning 400–700 nm at 10 nm intervals using a multilayer perceptron, with details of the calibration and reconstruction procedures provided in Methods. For a static scene under fixed illumination and sensor gain, the dark-subtracted measurements scale with exposure time within the linear, unsaturated response range. The dark-subtracted encoded measurements $\mathbf{Y}_i^{\text{norm}}$ acquired at different exposures can therefore be normalized to a common reference exposure as

$$\alpha_i = \frac{t_i}{t_{\text{ref}}} = 2^{\text{EV}_i}, \qquad \mathbf{Y}_i^{\text{norm}} = \frac{\mathbf{Y}_i^{\text{raw}}}{\alpha_i}, \quad i = 1,2,\ldots,N \tag{1}$$

Here, $N$ denotes the number of input exposures, $t_i$ is the exposure time, and $t_{\text{ref}}$ is the reference exposure time. The exposure value $\text{EV}_i$ denotes the exposure offset in stops relative to the reference, with $\text{EV} = 0$ corresponding to $t_{\text{ref}}$. In practice, noise, nonlinearity, and saturation introduce exposure- and channel-dependent deviations[31], motivating reliability-aware fusion.

As shown in Figure 1(b), the framework estimates measurement reliability from the dark-subtracted measurements, while fusion is performed after exposure normalization. Each raw exposure is independently processed by a shared reliability estimator comprising an analytical physical prior and a learned adjustment. A physical prior $\boldsymbol{P}_i \in \mathbb{R}^{C\times H\times W}$ with a hat-shaped profile tackles the fundamental physics of the sensor, which favors the reliable intermediate-response range and suppresses weak signals affected by noise and high signals approaching saturation or nonlinearity. To account for spatial and channel-dependent hardware deviations not captured by the theoretical prior, we employ a shared learned adjustment head $\boldsymbol{A}_i \in \mathbb{R}^{C\times H\times W}$ to predict adaptive corrections. These two components are combined to generate channel-specific fusion weights

$$\omega_i = \text{Softmax}_i[\log(\boldsymbol{P}_i + \tau) + \boldsymbol{A}_i], \tau = 0.01 \tag{2}$$

The inferred weights are applied to the exposure-normalized measurements to obtain the initial HDR estimate $\mathbf{Y}_{\text{fuse}}$. Exposure normalization places all measurements on a common scale, while channel-specific weighting models reliability independently across encoding channels.

The fused estimate is further corrected by a lightweight residual refinement module, which corrects residual errors to produce the final HDR encoded measurement $\mathbf{Y}_{\text{HDR}}$.

The shared estimator and symmetric aggregation make the framework invariant to exposure ordering and compatible with flexible numbers and configurations of input exposures.

The network is trained using a combination of $\mu$-law $L_1$ loss and measurement-domain spectral angle mapper (SAM) loss $\mathcal{L}_{\text{SAM}}^{\text{meas}}$ . The former balances reconstruction errors across the vast HDR signal range, while the latter constrains the relative response structure across encoding channels. The $\mu$-law transformation is used only for loss computation, leaving the fused HDR measurement in the linear domain. Detailed loss definitions and training settings are provided in Methods.

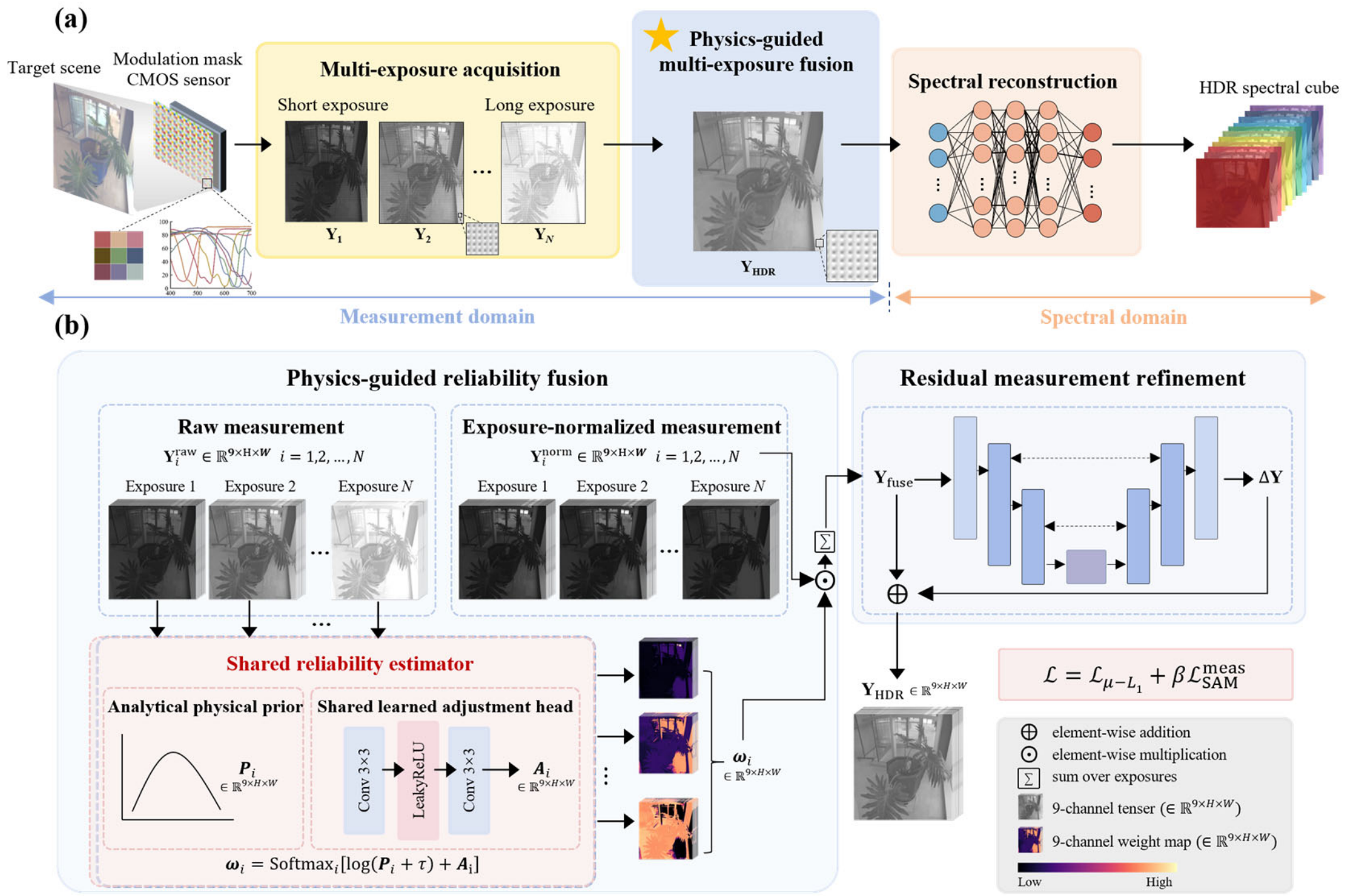

Figure 1. Physics-guided multi-exposure fusion for high-dynamic-range (HDR) computational hyperspectral imaging. (a) Imaging pipeline. The incident scene spectrum is encoded by the broadband modulation layer and image sensor into multi-channel measurements. Multi-exposure measurements are fused in the encoded measurement domain to obtain $\mathbf{Y}_{\text{HDR}}$, which is subsequently used to reconstruct a 31-band hyperspectral data cube spanning 400–700 nm. (b) Physics-guided multi-exposure fusion architecture. Each dark-subtracted raw exposure is independently processed by a shared reliability estimator comprising an analytical physical prior $\boldsymbol{P}_i$ and a learned adjustment $\boldsymbol{A}_i$ to generate spatially and channel-wise varying weights $\boldsymbol{\omega}_i$. The exposure-normalized measurements are symmetrically aggregated and refined to produce $\mathbf{Y}_{\text{HDR}}$. The shared formulation supports flexible numbers and configurations of input exposures. Training uses a combination of $\mu$-law $L_1$ loss and measurement-domain spectral angle mapper (SAM) loss $\mathcal{L}_{\text{SAM}}^{\text{meas}}$.

# 3. Results and Discussion

## 3.1 HDR Encoded-Measurement Fusion and Spectral Reconstruction

We evaluated the developed framework on HDR measurements synthesized from the ICVL[32] hyperspectral dataset using a forward model incorporating the experimentally measured spectral responses of nine encoding channels (Ch1–Ch9) described in Methods. Twenty independent test scenes under four three-exposure configurations were used for evaluation. Comparisons included analytical hat-weighted fusion and three representative learning-based HDR approaches: weight-map-based MEFNet[22], attention-guided AHDRNet[21], and transformer-based SCTNet[25]. Table 1 presents a controlled comparison under a unified training objective.

**Table 1.** Quantitative comparison of HDR measurement reconstruction, hyperspectral reconstruction, and computational efficiency on the ICVL-derived HDR test dataset.

| Method | HDR measurement domain | | | Spectral domain | | | Computational efficiency | | |
|---|---|---|---|---|---|---|---|---|---|
| | RMSE | SAM (rad) | PSNR (dB) | RMSE | SAM (rad) | PSNR (dB) | Parameters (M) | MACs (G) | Time (ms) |
| Ideal HDR GT | / | / | / | **0.00584** | **0.0286** | **44.968** | / | / | / |
| Single exposure | / | / | / | 0.02173 | 0.1752 | 34.119 | / | / | / |
| Hat | 0.00917 | 0.0182 | 46.220 | 0.01334 | 0.0614 | 37.542 | / | / | / |
| MEFNet | 0.01036 | 0.0191 | 44.729 | 0.01515 | 0.0664 | 36.496 | **0.026** | **1.29** | **1.28** |
| AHDRNet | 0.00330 | 0.0042 | 54.670 | 0.00664 | 0.0318 | 43.796 | 1.526 | 385.64 | 47.88 |
| SCTNet | 0.00366 | 0.0041 | 54.627 | 0.00650 | 0.0318 | 44.014 | 1.004 | 301.77 | 581.69 |
| Ours | **0.00169** | **0.0021** | **60.362** | **0.00609** | **0.0295** | **44.606** | 1.291 | 75.65 | 8.45 |

Note: Results are averaged over 20 ICVL test scenes under four three-exposure configurations: [-2, 0, +2], [-1, 0, +1], [-2, 0, +1.5], and [-2, +1, +1.5] EV. All learning methods are adapted to 9-channel encoded measurements and optimized using the same $\mu$-law $L_1$ loss and measurement-domain angular supervision. Metrics are reported for 9-channel HDR measurements and reconstructed 31-band spectra. Measurement-domain errors were evaluated on linear encoded measurements, whereas spectral-domain RMSE and PSNR were calculated after L2 normalization **(Supplementary Note S1)**. Single exposure denotes the EV = 0 input. Time and MACs are measured on one NVIDIA GeForce RTX 5080 GPU (16 GB) under 512 × 512 resolution. Abbreviations: GT, ground truth; RMSE, root-mean-square error; SAM, spectral angle mapper; PSNR, peak signal-to-noise ratio; MACs, multiply–accumulate operations.

As shown in Table 1, our method reduces measurement-domain root-mean-square error (RMSE) and SAM by approximately 49% relative to the best competing learning-based methods. These gains extend to the spectral domain, reducing RMSE and SAM by 6.3% and 7.2%, respectively, and improving PSNR by 0.592 dB over SCTNet. Although spectral reconstruction remains imperfect even with ideal HDR measurements, our method closes approximately 62% of the spectral RMSE gap between SCTNet and this ideal-input reference. The competing methods generally achieve lower spectral accuracy under their original training objectives (Supplementary

Table S1). Our method also retains a favorable accuracy–efficiency balance, running 5.7× faster than AHDRNet and 68.8× faster than SCTNet.

Figure 2(a) illustrates the complementary information provided by different exposures. Short exposures preserve highlight details but suffer from noise in dark regions, whereas long exposures recover weak signals at the expense of highlight saturation. Our fusion preserves information from both regions, as further demonstrated by the reconstructed spectra in Figure 2(b).

Figure 2(c) reveals distinct exposure preferences across encoding channels, consistent with their unequal spectral responsivities and signal-dependent reliability. Compared with the smooth and fixed hat weighting, the learned adjustment produces more distinct transitions between regions favoring different exposures (Supplementary Figure S1). At the dark location P2, our method assigns greater weight to the long exposure for most channels while retaining channel-dependent preferences, compared with the more distributed hat weights **(**Supplementary Figure S2**)**. These observations illustrate how the learned adjustment refines the physical reliability prior to accommodate differences among encoding channels. Compared with other HDR fusion methods, our method yields lower spectral-angle errors, particularly in regions affected by exposure-dependent degradation [Figure 2(d)].

Ablation studies in Supplementary Table S2 confirm that explicit reliability-aware fusion followed by residual refinement outperforms direct reconstruction from concatenated exposures. Combining the analytical physical prior with learned adjustment improves spectral reconstruction over either component alone, while channel-specific weighting increases PSNR from 44.180 to 44.606 dB compared with channel-shared weighting. The shared reliability estimator also accommodates varying numbers and configurations of input exposures without architectural modification or retraining (Supplementary Figure S3). With a single short-exposure input, the framework naturally performs denoising and restoration, achieving a spectral PSNR of 41.609 dB and a SAM of 0.0396 rad. Supplementary Table S3 further shows that measurement-domain angular supervision consistently improves spectral reconstruction across architectures, supporting the importance of preserving inter-channel relationships for subsequent spectral inversion.

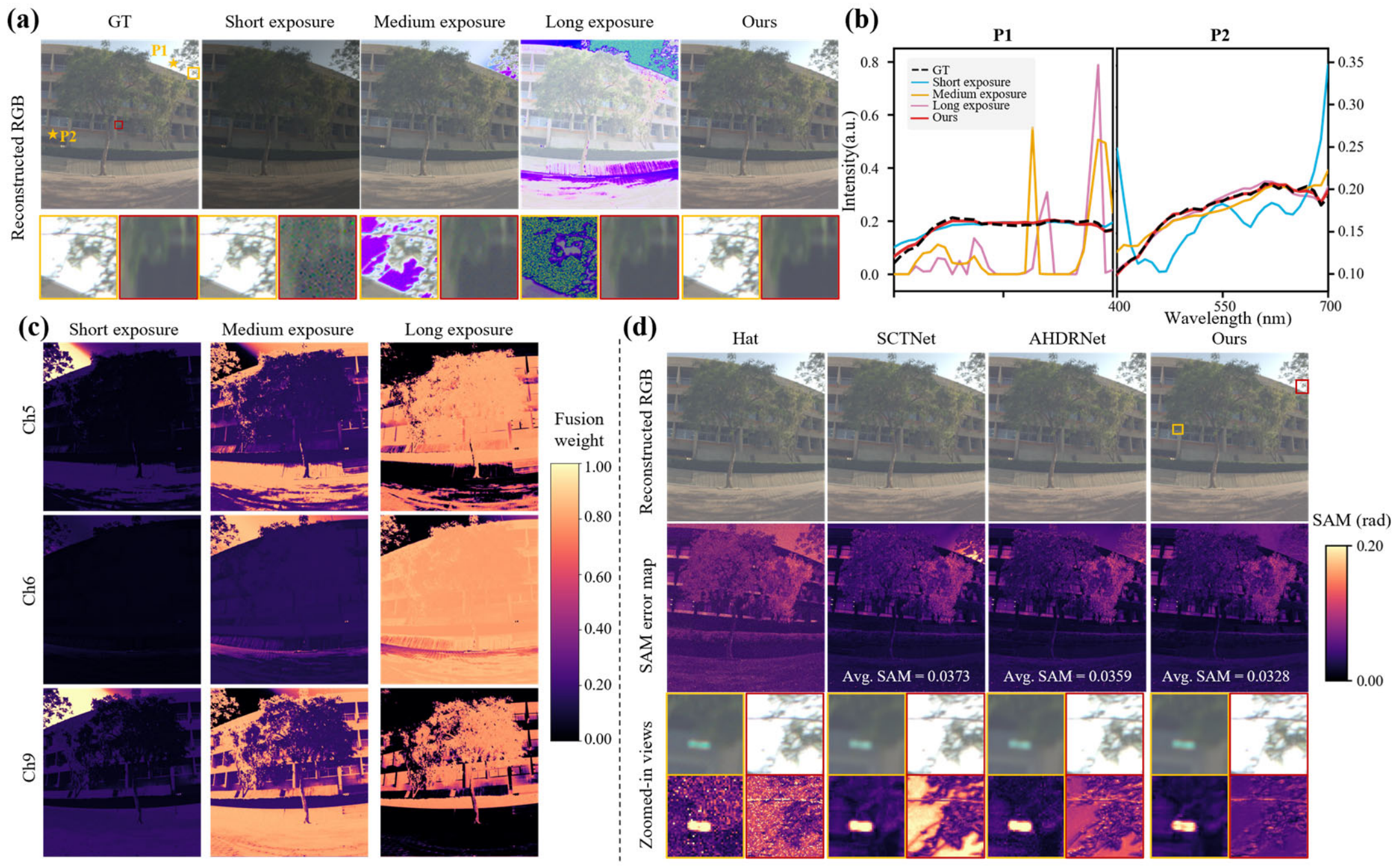

**Figure 2. Physics-guided HDR measurement fusion and hyperspectral reconstruction.** (a) RGB renderings of the ground truth (GT), three single-exposure reconstructions at -2, 0, and +2 EV, and our HDR result. GT is reconstructed from ideal HDR encoded measurements using the same spectral reconstruction network. Yellow and red boxes indicate enlarged regions. (b) Reconstructed spectra at representative highlight (P1) and dark (P2) locations compared with GT. (c) Spatial fusion-weight maps of three encoding channels (Ch5, Ch6, and Ch9) across the three exposures, illustrating exposure- and channel-dependent weighting. All maps share a color scale from 0 to 1. (d) Comparison with representative HDR fusion methods. RGB visualizations, pixel-wise spectral angle mapper (SAM) error maps, and enlarged regions are shown for the scene in (a), with the average SAM reported for each method. The enlarged RGB regions are independently exposure-stretched for visualization.

## 3.2 Sensor-Response Characterization and Experimental Spectral Validation

The fusion formulation assumes that exposure-normalized measurements remain reliable within the linear-response range. In practice, individual CMOS encoding channels may deviate from linearity before nominal saturation[33]. We therefore characterized the reliable response range of each channel by sampling measurements across the entire effective field of view, rather than restricting the analysis to a small central region. This full-field approach accounts for spatial variations caused by lens shading and nonuniform illumination, with the measurements aggregated for channel-wise analysis (Supplementary Note S2).

Figure 3(a) reveals distinct response behaviors among the nine channels. Most remain approximately linear over a broad signal range, whereas individual channels can deviate before reaching nominal saturation, as observed for Ch6. Channel-wise lookup table (LUT) correction reduces systematic nonlinearity but residual variations remain substantial at high signal levels [Supplementary Figure S4]. We therefore introduced a channel-aware reliability gate [Figure 3(b)] to suppress measurements beyond the characterized reliable range, reducing localized reconstruction artifacts (Supplementary Figure S5).

We further validated the system using a 24-patch ColorChecker, comparing the reconstructed spectra against independent spectrometer measurements. Comparisons included analytical hat fusion and SCTNet. For SCTNet, unreliable measurements were replaced with estimates from a reliable exposure before inference, whereas our framework incorporates sensor-specific reliability constraints directly into the fusion stage. As shown in Figure 3(c), our method suppresses noise and fixed-pattern artifacts while preserving color fidelity, whereas SCTNet exhibits a noticeable color bias. At patch P1 [Figure 3(d)], our reconstructed spectrum closely follows the spectrometer reference, achieving the lowest SAM of 0.0285 rad.

Evaluation over all 24 ColorChecker patches further confirms this trend [Figure 3(e)]. Our method reaches 38.341 and 38.992 dB with two and three exposures, respectively, exceeding the best single-exposure result of 35.792 dB. In comparison, analytical hat fusion reaches 35.252 and 36.189 dB under the same configurations, indicating greater sensitivity of fixed pointwise weighting to noisy or locally unreliable measurements when exposure redundancy is limited.

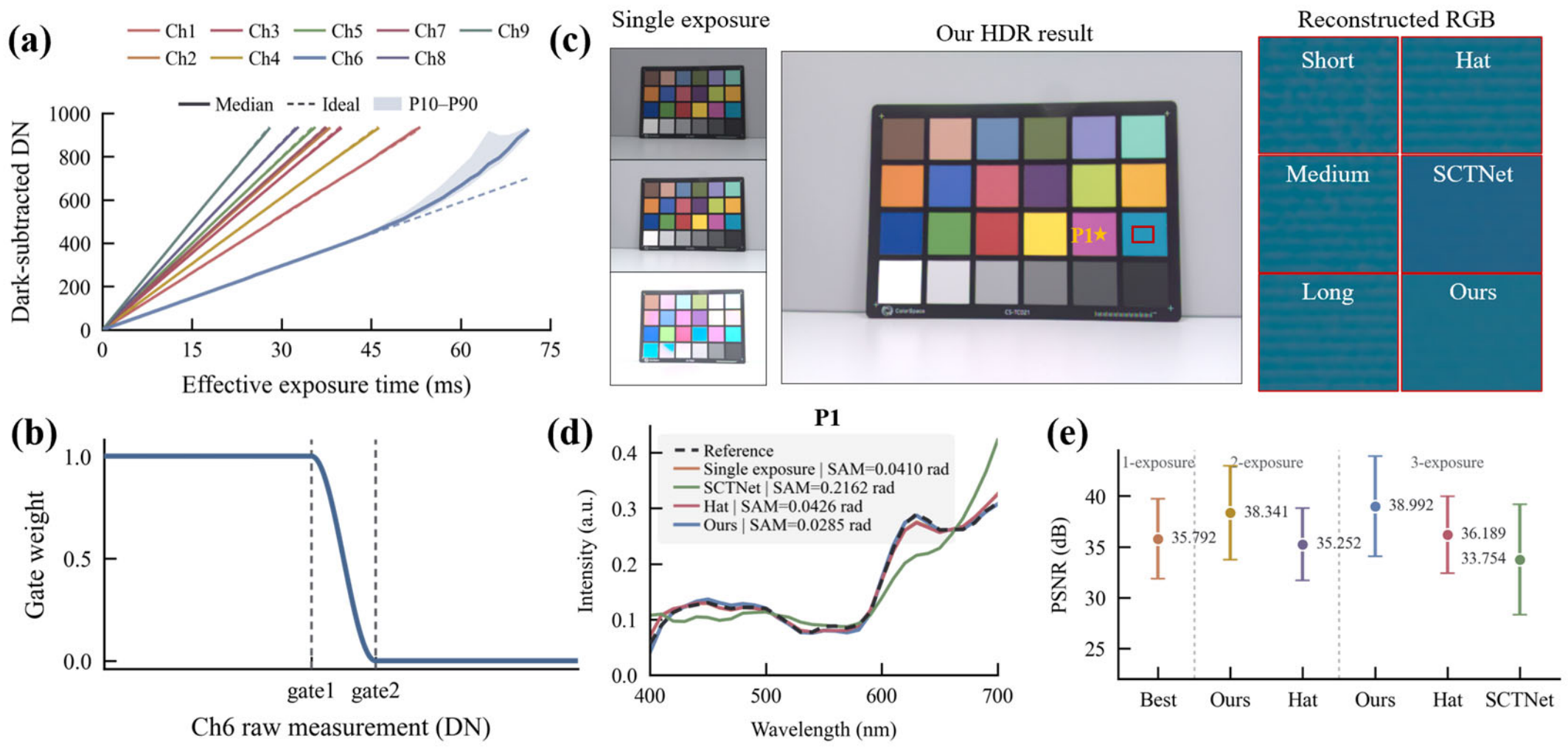

**Figure 3. Full-field sensor-response characterization and quantitative real-system validation.** (a) Full-field responses of the nine encoding channels. Solid curves and shaded regions denote the median and P10–P90 distributions within equivalent-exposure bins, respectively; dashed lines indicate ideal linear responses fitted in the low-signal regime. Increased dispersion at high responses indicates reduced measurement confidence. (b) Channel-aware reliability gate derived from the characterized reliable response range. (c) Multi-exposure ColorChecker imaging acquired at 7.8, 15.6, and 62.5 ms, with reconstructed RGB results and enlarged comparisons of individual exposures, hat fusion, SCTNet, and our method. (d) Reconstructed spectrum at P1 compared with the spectrometer reference (OceanOptics, ST-VIS-25); SAM values are reported in the legend. (e) Spectral reconstruction performance over 24 ColorChecker patches. Pointwise reference spectra were spatially replicated within each patch for evaluation. Markers and error bars denote the mean

and standard deviation of patch-wise spectral PSNR, respectively. DN denotes digital number.

### 3.3 Quantitative Dynamic-Range Evaluation and Real-Scene Hyperspectral Imaging

To evaluate HDR hyperspectral imaging over a broad range of incident light intensities, we imaged the 20 peripheral optical-density (OD) patches of an OECF test chart (TE264, Image Engineering) under illuminant A at exposure times of 1, 5, and 62.5 ms [Figure 4(a)]. The patches span an OD range of 0.02–4.19. The signal from the darkest patch (OD = 4.19) was too weak for reliable spectrometer measurements and was therefore excluded from quantitative analysis. The reconstructed RGB images illustrate the complementary information provided by different exposures, with shorter exposures preserving highlights and longer exposures revealing darker regions. Our fusion combines these measurements while suppressing visible artifacts.

We evaluated spectral reconstruction performance using spectral-domain PSNR against independently measured reference spectra for the remaining 19 patches [Figure 4(b)]. We define the spectral reconstruction dynamic range (SRDR) as the range of incident light intensities over which high-fidelity spectral reconstruction is maintained. A spectral-domain PSNR threshold of 30 dB was used to determine this range, as described in Supplementary Note S3. Our method achieved an SRDR of 42.8 dB, exceeding the best single exposure (26.8 dB) by 16.0 dB. Analytical hat-weighted fusion achieved 38.8 dB, whereas our method preserved higher spectral fidelity in high-transmittance regions. At low transmittances, both fusion methods performed similarly and eventually fell below the longest exposure, suggesting that low signal-to-noise ratios may limit reliable fusion. The SRDR improvement over the best single exposure persisted across PSNR thresholds of 28–31 dB (Supplementary Table S4). As the brightest patch still met the fidelity criterion, the measured SRDR is bounded by the tested OD range.

We further validated the system in two representative HDR scenes [Figure 4(c)]. Scene 1 was constructed in a D65-illuminated light box (ColorSpace, CS-LE006-QC), with localized broadband illumination on a ColorChecker, a reflective white board, and a dark enclosure. Scene 2 combined indoor objects with an exterior window view containing a sunlit wall and sky. In both scenes, intermediate exposures captured most regions but produced pronounced color artifacts in highlights. Our fusion suppressed these artifacts and low-signal noise while preserving spatial details. Spectral comparisons at P1–P4 in Figure 4(d) further demonstrated the improvement. The fused spectra closely matched the spectrometer references despite substantial deviations in the medium- and long-exposure measurements at P2–P4 and the long-exposure measurement at P1.

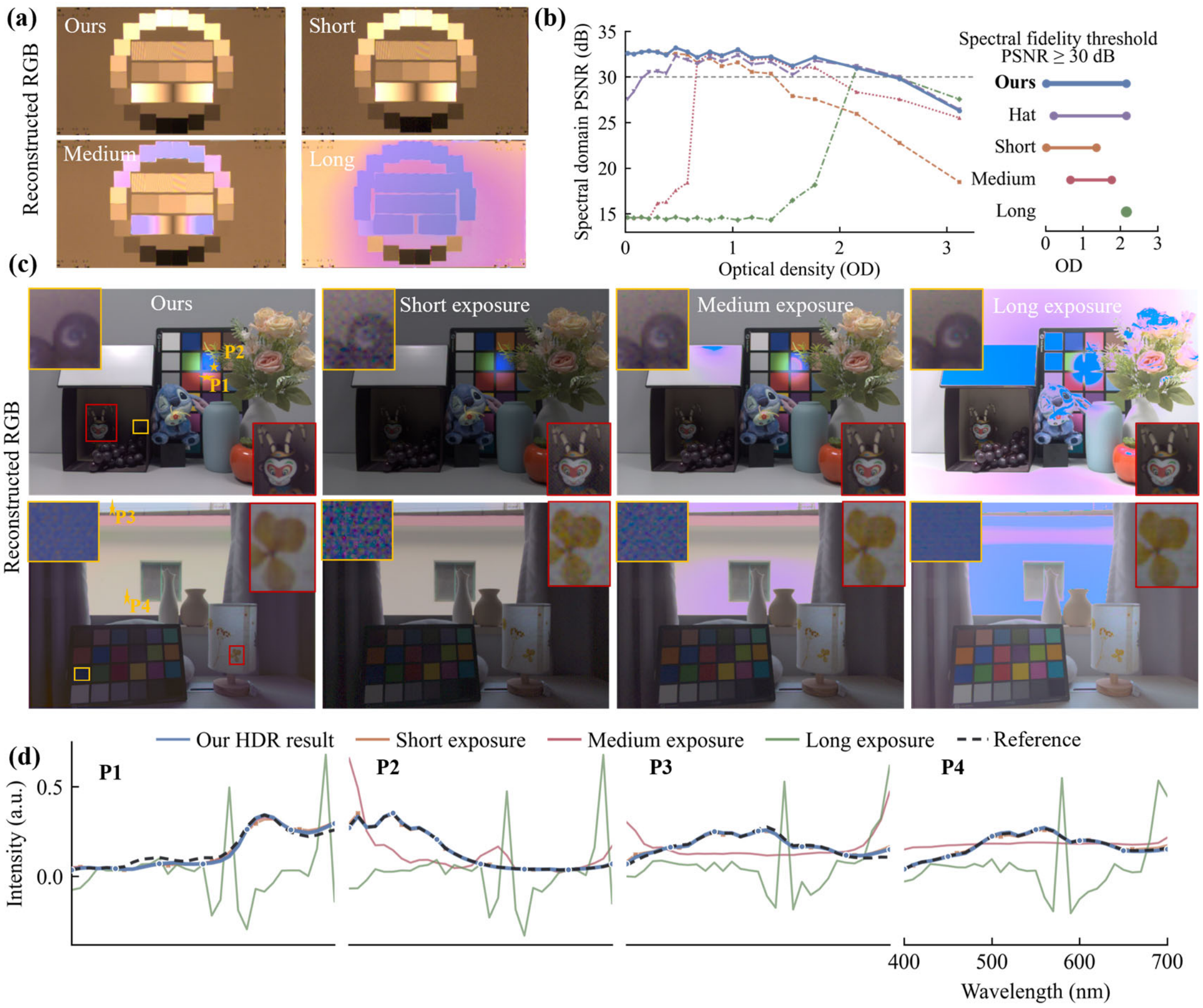

**Figure 4. Quantitative dynamic-range evaluation and hyperspectral imaging of real HDR scenes.** (a) Reconstructed RGB visualizations synthesized from hyperspectral results for a 20-patch OECF test chart (TE264, Image Engineering) under illuminant A, obtained from three individual exposures (1, 5, and 62.5 ms) and our HDR fusion. The peripheral patches span an optical density (OD) range of 0.02–4.19. (b) Spectral-domain PSNR across the OD patches for individual exposures, analytical hat-weighted fusion, and our method, evaluated against independently measured reference spectra. The dashed line indicates the 30 dB spectral fidelity criterion, with colored bars showing the corresponding usable OD ranges. (c) Reconstructed RGB visualizations synthesized from hyperspectral results for two representative HDR scenes under three individual exposures and our HDR fusion. Exposure times were 3.9, 15.6, and 62.5 ms (Scene 1) and 2, 5, and 31.2 ms (Scene 2). Yellow and red boxes indicate enlarged regions. The same μ-law mapping with a display parameter of 60 was applied to all RGB results of Scene 2. (d) Reconstructed spectra at P1–P4 marked in (c); black dashed lines denote the reference spectra measured by a commercial spectrometer.

## 4. Conclusion

In summary, this work extends the usable dynamic range of computational hyperspectral imaging by addressing measurement reliability before spectral inversion. Unlike conventional HDR fusion methods that primarily target final display images, our framework combines complementary exposures directly in the encoded measurement domain, using an analytical reliability prior and learned channel-specific adjustments to preserve the inter-channel relationships required for spectral reconstruction. The set-wise and modular formulation further supports flexible exposure configurations and integration with existing hardware calibration pipelines. Synthetic and experimental results demonstrate improved measurement recovery and spectral fidelity. To quantify the range supporting high-fidelity spectral reconstruction, we introduced the spectral reconstruction dynamic range (SRDR). Measurements with an optical-density target demonstrate a 16.0 dB increase in SRDR over the best fixed single exposure, complemented by spectral validation in real HDR scenes.

While the current framework assumes static, spatially aligned scenes and does not address inter-exposure motion, its software-level approach extends the usable measurement range without modifying the sensor hardware. Future work will explore motion compensation and co-design with adaptive optical modulation hardware, such as digital micromirror devices (DMDs), to enable acquisition and reconstruction to adapt jointly to scene conditions. Such integration could advance computational spectral imaging from passive measurement toward adaptive sensing.

# 5. Methods

## 5.1 Computational Hyperspectral Imaging System

The snapshot 9-channel hyperspectral camera used in this study integrates a broadband thin-film modulation array with a CMOS image sensor. The modulation layer consists of a cured colloidal thin film patterned by photolithography, with nine distinct encoding channels arranged in a periodic 3 × 3 mosaic and aligned one-to-one with the underlying sensor pixels. The sensor has a pixel size of 1.75 × 1.75 $\mu m^2$, a raw resolution of 1,600 × 1,200 pixels, and a 1/5-inch sensor format. A photograph of the camera prototype and an optical micrograph of the on-chip array are provided in Supplementary Figure S6. To construct the 9-channel measurements, a 3 × 3 sliding window moves across the raw mosaic image with a stride of one pixel, collecting one measurement from each channel at every position. Following edge cropping, the reconstruction yields a hyperspectral data cube with output spatial dimensions of 1,500 × 1,080 pixels.

An external visible-band filter limits the camera's effective spectral sensitivity to the operational 400–700 nm range. The effective spectral responses of all nine channels were characterized using a customized characterization system. The corresponding response profiles and correlation analysis are provided in Supplementary Figure S7.

Under the local spectral approximation, the nine samples collected by each sliding window are treated as measurements of the same scene spectrum. At spatial location $(x,y)$, the ideal response of channel $c$ at reference exposure $t_{\mathrm{ref}}$ is

$$y_{\mathrm{ref},c}^{\mathrm{ideal}}(x,y)=t_{\mathrm{ref}}\int_{\lambda_{\min}}^{\lambda_{\max}} s(x,y,\lambda)\, r_c(\lambda)\mathrm{d}\lambda,\ c = 1,2,\dots,C, \tag{3}$$

where $s(x,y,\lambda)$ denotes the incident spectrum and $r_c(\lambda)$ represents the spectral response of channel $c$.

## 5.2 Hyperspectral Reconstruction Pipeline

Following dark-level subtraction, multi-exposure measurements were corrected using the channel-specific response LUT and the characterized channel-aware reliability gate. These measurements were exposure-normalized and fused in the encoded measurement domain. The resulting HDR measurement underwent channel-wise flat-field correction to compensate for spatial response nonuniformity and was calibrated using standard reference spectra. Final spectral reconstruction was performed using a pixel-wise multilayer perceptron (MLP) featuring a 9-500-500-500-31 architecture with LeakyReLU activations, which maps the 9-dimensional encoded vector to a 31-dimensional output spanning 400–700 nm at 10 nm intervals. The same reconstruction network was used throughout this work for downstream spectral reconstruction. Detailed dataset construction and training settings are provided in Supplementary Note S4.

### 5.3 HDR Dataset Generation and Network Training

The training dataset was synthesized from the ICVL hyperspectral dataset. Hyperspectral cubes were projected into the measurement domain using the measured spectral responses of the nine encoding channels according to Eq. (3). Unclipped encoded measurements served as HDR ground truth, while low-dynamic-range (LDR) measurements were generated through exposure scaling, Poisson–Gaussian noise simulation, clipping, and 10-bit quantization to mirror the experimental camera. The dataset comprised 160 training, 22 validation, and 20 independent test scenes.

The fusion network employs a shared two-layer convolutional adjustment head with a 9–16–9 channel configuration. The learned channel-specific adjustments are combined with the analytical prior to determine the exposure weights, followed by a three-scale residual U-Net to refine the fused measurements. The network was trained for 500 epochs using a combination of $\mu$-law $L_1$ and measurement-domain SAM losses, with the Adam optimizer implemented in PyTorch on an NVIDIA GeForce RTX 5080 GPU. Further details of data synthesis, exposure configurations, network architecture, and training are provided in Supplementary Note S5.

ASSOCIATED CONTENT

**Supporting Information**

Additional quantitative comparisons and ablation studies; analysis of fusion weights and generalization across exposure configurations; sensor-response characterization, nonlinearity correction, and reliability gating; camera hardware and spectral-response characterization; and detailed evaluation metrics, dynamic-range analysis, dataset synthesis, network architectures, and training procedures (PDF).

`